\documentclass{LT23auth}
\usepackage[dvips]{graphicx}

\begin{document}

\begin{frontmatter}

\title{The effect of large quantum fluctuation on 
the noise of a single-electron transistor}

\author[address1]{Yasuhiro Utsumi\thanksref{thank1}},
\author[address1]{Hiroshi Imamura},
\author[address1]{Masahiko Hayashi},
\author[address1]{Hiromichi Ebisawa}

\address[address1]{Graduate School of Information Sciences, 
Tohoku University, Sendai 980-8579, Japan}

\thanks[thank1]{Present address: 
Max-Planck-Institut f\"ur Mik\-rostruk\-turphy\-sik
Weinberg 2, 
Halle, Germany. E-mail:utsumi@mpi-halle.de}

\begin{abstract}
We theoretically investigate the noise of a single-electron transistor 
in the regime of large quantum fluctuation of charge out of equilibrium. 
We show that the charge noise is suppressed due to the charge renormalization 
caused by the quantum fluctuation. 
However the fluctuation is not strong enough to wash out 
the charge quantization. 
We find that the renormalization effect reduces the performance of 
a single-electron electrometer. 
\end{abstract}

%
%
\end{frontmatter}

\newcommand{\rd}{{\rm d}}
\newcommand{\ri}{{\rm i}}
\newcommand{\mtau}{\mbox{\boldmath$\tau$}}
\newcommand{\mat}[1]{\mbox{\boldmath$#1$}}

%
%
In a small metallic island where the charging energy $E_C$ exceeds 
the temperature $T$ (we use the unit $k_{\rm B}=1$), 
the Coulomb interaction affects transport properties. 
The resulting phenomenon called the Coulomb blockade 
has attracted much attention in the last decade. 
Especially, the quantum fluctuation of charge in a single-electron 
transistor (SET) is one of the basic problems in this field. 
The quantum fluctuation is quantitatively characterized by a parameter, 
dimensionless conductance
$\alpha_0=R_{\rm K}/((2 \pi)^2 R_{\rm T})$, 
where  $R_{\rm K}=h/e^2$ is the quantum resistance 
and $R_{\rm T}$ is the parallel tunneling resistance of 
the source and the drain junction. 
Recently, there has been much development in theoretical investigations
in the whole range of the tunneling strength. 
In the {\it weak tunneling regime}, where 
the life-time broadening of a charge state level is much smaller than
the typical level spacing of charge states, $\alpha_0<1$, 
the lowest two charge states well describe the low-energy physics. 
At the degeneracy point where the energy difference between 
two charge states $\Delta_0 \propto E_C$ is zero, 
the charge number fluctuates greatly. 
At this point, the conductance and charging energy renormalization 
below the Kondo temperature
$T_{\rm K}=E_C / (2 \pi) {\rm e}^{-1/(2 \alpha_0)}$ 
has been predicted\cite{Schoeller}. 

%
%
Though previous investigations have revealed much about 
the quantum fluctuation, 
they are limited to the discussion on average values. 
In order to understand quantum nature of the charge fluctuation, 
investigations on the statistical property of carriers, i.e. the noise, 
are required. 
The noise in the weak tunneling regime is also important 
for practical applications, because it determines the 
performance of SET electrometers
\cite{KorotkovR,Johansson}. 


%
In this paper, we discuss the charge noise and the energy sensitivity. 
They are calculated based on the Schwinger-Keldysh approach and 
the drone (or Majorana) fermion representation\cite{Utsumi,Isawa}. 
We reformulated and extended the resonant tunneling 
approximation\cite{Schoeller} to the noise expression in a charge 
conserving way. 
In a previous paper\cite{Utsumi}, we discussed the region 
$eV \gg T_{\rm K}$ where the renormalization effect is unimportant. 
In this paper we discuss the opposite region $eV< T_{\rm K}$ and 0 K
using the expression in Ref. \cite{Utsumi}. 

\begin{figure}[h]
\begin{center}\leavevmode
\includegraphics[width=0.78\linewidth]{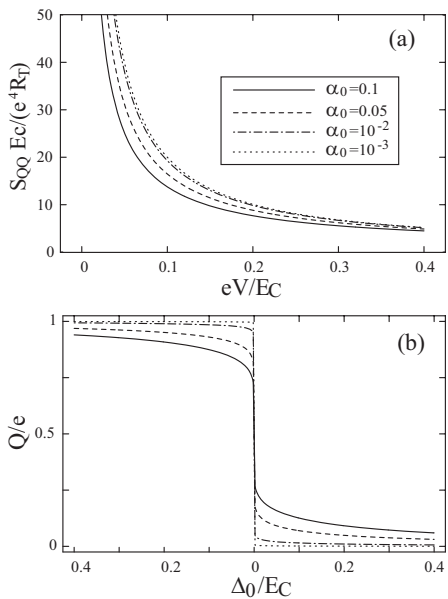}
\caption{
(a) The bias voltage dependence of the charge noise 
at $\Delta_0=0$ and 0 K. 
(b) The average charge as a function of $\Delta_0$ at $eV=0$. 
Solid, dashed, dot-dashed and dotted curves show results for 
$\alpha_0=0.1$, $0.05$, $10^{-2}$ and $10^{-3}$, respectively.
}
\label{fig:SQQ}\end{center}\end{figure}

%
%
Figure \ref{fig:SQQ} (a) shows the bias voltage dependence of the 
normalized charge noise. 
As $\alpha_0$ increases, it decreases because of the 
charge renormalization. 
The charge renormalization means that the charge noise becomes
the same form as the orthodox theory with renormalized parameters, 
the renormalized conductance $\tilde{\alpha_0}=z \alpha$, 
charging energy $\tilde{\Delta_0}=z \Delta_0$
and 
charge $\tilde{e}=z e$: 
$4 \tilde{e}^2 \tilde{\Gamma}_{+}
\tilde{\Gamma}_{-}/ (\tilde{\Gamma}_{+}+\tilde{\Gamma}_{-})^3$. 
The transition rate of a electron tunneling into 
(out of) the island 
$\tilde{\Gamma}_{+(-)}$ is estimated by the golden rule. 
The renormalization factor is given by $z=1/(1+2 \alpha_0 \ln(E_C/(|eV|/2)))$. 
It is noticed that the result is inconsistent with a naive expectation;
from the Johnson-Nyquist formula 
$S_{QQ} = 4 R_{\rm T} \, T \, C^2$, 
the charge noise
is expected to be proportional to $z^{-3}$. 

Though the normalized charge noise is suppressed, 
it diverges at $eV=\Delta_0=0$. 
It means that the charge changes by \lq \lq one" at 
the degeneracy point even at the large quantum fluctuation. 
This is confirmed by the fact that the 
slope of excitation energy dependence of the average charge 
diverges at the degeneracy point (Fig. \ref{fig:SQQ} (b)). 

%
%
\begin{figure}[h]
\begin{center}\leavevmode
\includegraphics[width=0.78\linewidth]{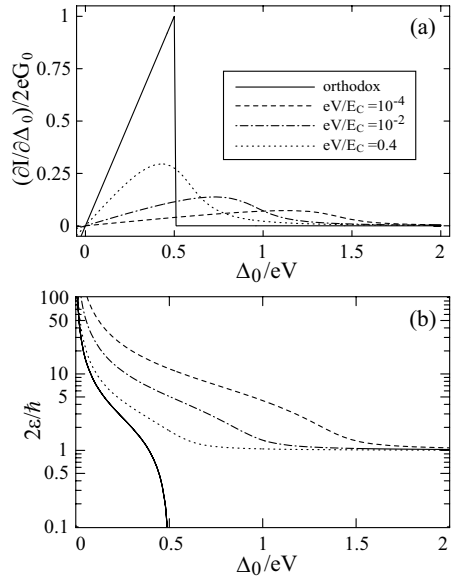}
\caption{
The excitation energy dependence of 
(a) the slope of the average current and
(b) the energy sensitivity for $\alpha_0=0.1$ at
$eV/E_C=10^{-4}$ (dashed curve), 
$10^{-2}$ (dot-dashed curve) and 
$0.4$ (dotted curve) at 0 K.
Solid curves show results of the orthodox theory. 
$G_0=1/(R_{\rm L}+R_{\rm R})$ is the series conductance. 
}
\label{fig:ense}\end{center}\end{figure}

We discuss the renormalization effect on the performance of a 
SET electrometer. 
Figures \ref{fig:ense} (a) and (b) show the excitation energy 
dependence of the slope of $I$-$V$ characteristic and the energy sensitivity 
for various bias voltage with large $\alpha_0$. 
As the bias voltage decreases, the peak shifts rightwards
because of the charging energy renormalization\cite{Schoeller}. 
Moreover the peak hight is strongly suppressed. 
This is because both the charging energy and the conductance are renormalized, 
the slope $\partial I/\partial \Delta_0$ is approximately
proportional to $z^2$. 
As $S_{II} \propto z$ and $S_{QQ} \propto z$, 
the energy sensitivity 
$(\hbar/2) \sqrt{S_{QQ} S_{II}}/|\partial I / \partial \Delta_0|$
is inversely proportional to $z$. 
This fact means that the renormalization effect reduces the 
energy sensitivity as shown in Fig. \ref{fig:ense} (b). 

%
%
In conclusion, we investigated effect of the renormalization 
on the noise theoretically. 
We showed that the charge noise is suppressed due to 
the charge renormalization. 
However the quantum fluctuation does not wash out the charge quantization
for arbitrary $\alpha_0$ in the weak tunneling regime. 
The renormalization effect reduces the performance of SET electrometer because 
both the conductance and the charging energy renormalization tend to 
suppress the slope of $I$-$V$ characteristic. 

%
%
We would like to thank Y. Isawa, J. Martinek, Yu. V. Nazarov and G. Johansson
for variable discussions and comments. 
This work was supported by a Grant-in-Aid for Scientific
Research (C), No. 14540321 from MEXT. H.I. was supported by MEXT,
Grant-in-Aid for Encouragement of Young Scientists, No.
13740197.

%
%

\end{document}